# Inverse Design of Realizable Metasurface based Absorbers using Improved Conditioning and Diversity Enhanced Progressively Growing GANs

Vineetha Joy[1*], Mohammad Abdullah[1], Pramit Pal[2], Anshuman Kumar[3], Amit Sethi[3], Hema Singh[1]
[1]Centre for Electromagnetics, CSIR-National Aerospace Laboratories, Kodihalli, Bangalore 560017
[2]Birla Institute of Technology and Science, Pilani, Rajasthan 333031
[3]Indian Institute of Technology, Bombay, Maharashtra 400076
vineethajoy@nal.res.in, pramit.pal2005@gmail.com; abdullahm16367@gmail.com; anshuman.kumar@iitb.ac.in, asethi@iitb.ac.in, hemasingh@nal.res.in

***Abstract*— Metasurfaces enable precise manipulation of electromagnetic (EM) waves for applications such as beam steering, sensing, and stealth technology. However, inverse design of metasurfaces with targeted EM responses remains challenging due to the computational expense of iterative full-wave simulation-driven optimization and the limited conditioning fidelity and diversity of existing generative approaches. To address these challenges, this paper presents a generative inverse design framework for controllable and physically consistent metasurface synthesis under continuous spectral constraints. The proposed approach employs a progressively growing Wasserstein generative adversarial network with gradient penalty (WGAN-GP) integrated with feature-wise linear modulation (FiLM)-based conditioning for stable propagation of continuous spectral and fabrication constraints. EM consistency is embedded directly into the generative learning process through a surrogate-assisted spectral alignment loss, enabling physics-constrained generation during training. Further, a determinantal point process (DPP)-based diversity regularization strategy is incorporated to generate geometrically diverse yet spectrally consistent realizations for the same target response. The effectiveness of the proposed framework is demonstrated through the generation of practically realizable metasurface absorbers exhibiting diverse reflection characteristics in the frequency range of 2–18 GHz. EM simulations validate that the generated designs meet the target specifications with high accuracy. The final proposed framework achieved an average mean squared error of 0.0052, accumulated average error of 0.0343, diversity score of 0.8730, band alignment accuracy of 0.8533, and a valid EM design generation rate of 89.57%, clearly demonstrating its capability to generate highly accurate, diverse, electromagnetically consistent and fabrication realizable metasurface configurations.**

## I. Introduction

Metasurfaces are two dimensional arrays of sub-wavelength scattering elements that provide unprecedented control over incident electromagnetic (EM) waves [1], [2]. They are the planar counterparts of bulk metamaterials. While metamaterials introduced attractive concepts such as negative refraction [3], [4], their bulky volumetric structure imposes significant fabrication and space constraints. In contrast, metasurfaces, composed of thin layers of engineered meta-atoms, impart spatially varying amplitude, phase and polarization modulation to incident waves [1], [5] by leveraging generalized Snell's law of refraction and reflection [6]. For instance, consider a plane wave striking a metasurface situated at the boundary between two media, Medium#I and Medium#II, with refractive indices $n_I$ and $n_{II}$ respectively as shown in Fig. 1. The generalized Snell's laws of refraction and reflection applicable to a uniform phase gradient along the boundary are expressed as [6], [7],

$$n_{II}\sin\theta_{II} - n_I\sin\theta_I = \frac{\lambda}{2\pi}\frac{d\Phi(x)}{dx} \quad (1)$$

$$\sin\theta_{ref} - \sin\theta_I = \frac{\lambda}{2\pi n_I}\frac{d\Phi(x)}{dx} \quad (2)$$

where $\lambda$, $\theta_I$, $\theta_{ref}$, $\theta_{II}$ and $d\phi/dx$ denote the free space wavelength, angle of incidence, angle of reflection, angle of refraction and phase gradient respectively. Equation (1) and Equation (2) show that the direction of incident wave and the shape of the wavefront can be modified by varying the spatial phase on a metasurface. Hence, metasurfaces enable focusing, beam steering, and holography within subwavelength thicknesses [8], [9] with added advantages of planar fabrication, miniaturization, and ease of electronic integration. Due to these attractive attributes, metasurfaces have been used for different applications like metalenses [10], [11], polarization control [12], [13], wavefront shaping [9], advanced sensing [14], [15], communication systems [13] and stealth technologies [16].

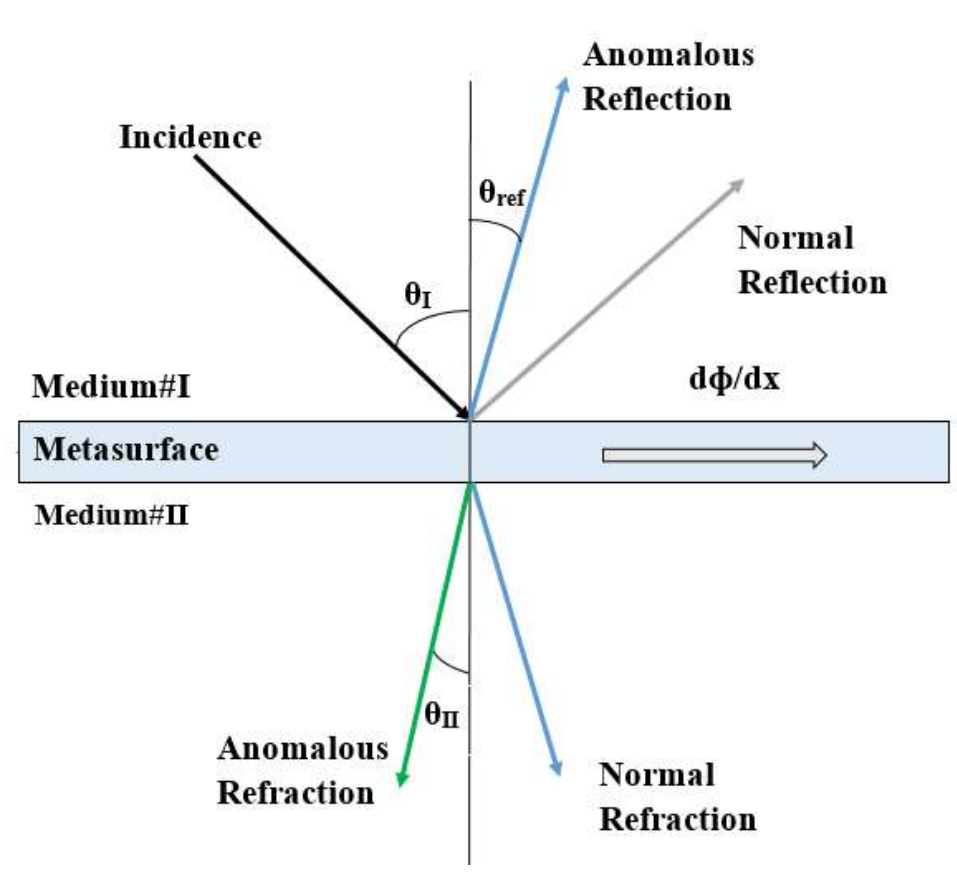

Fig. 1. EM wave incident on a metasurface.

The conventional methodology for the design of metasurfaces typically begins with an initial meta-atom geometry derived from underlying physical principles, followed by iterative optimization of its structural parameters. This process involves repeated adjustments and refinements through extensive parameter sweeps and full-wave EM simulations until the desired performance is achieved. However, such a brute-force approach becomes computationally prohibitive and increasingly unmanageable as the design space expands, particularly in scenarios involving high degrees of freedom [17], [18]. To overcome these constraints, global optimization algorithms such as genetic algorithms (GAs), particle swarm optimization (PSO), and differential evolution have been widely used in the design of metasurfaces [19]. Although these methods enable exploration of large design spaces, they suffer from high computational costs, convergence instability, and sensitivity to hyper parameter tuning [20], [21]. Furthermore, they often yield local optima and do not scale well with increasing structural complexity or tighter design constraints [21].

Recent advances in machine learning, particularly deep learning (DL), offer data-driven alternatives to conventional design approaches. Supervised DL models, including convolutional neural networks (CNNs) and multilayer perceptron (MLP), have been applied to learn mappings between target EM response and geometries [22], [23]. While such models significantly reduce inference time after training, they often require large labelled datasets and suffer from generalization limitations especially for the design of free-form shapes where high randomness and weak symmetry make it hard for conventional neural networks to generalize well.

To overcome the above limitations, researchers have turned to generative models like variational autoencoders (VAE), generative adversarial networks (GANs) and diffusion models, which can capture the full distribution of viable designs. Instead of learning a single deterministic mapping, generative models learn the underlying distribution of metasurface patterns that achieve a desired functionality. For example, a conditional GAN can be trained on a library of known metasurface designs and their responses, and thereafter generate many new design candidates that meet a target response [24]- [27]. Likewise, VAEs have been employed to embed metasurface geometries into a continuous latent space, allowing efficient search and interpolation for optimized designs [28]- [31]. By sampling the latent space, one can obtain multiple distinct solutions that satisfy the design goals. Further, diffusion models generate new designs by progressively denoising random inputs toward target-consistent outputs [32].

However, it is important to note that most generative models reported in the literature exhibit inherent limitations when applied to the inverse design of metasurfaces. In particular, VAEs often produce blurred reconstructions which are difficult to fabricate. Moreover, achieving accurate target responses typically needs additional optimization within the latent space which in turn introduces significant computational overhead. Further, existing GAN-based frameworks are constrained by mode collapse, convergence instability, and limited adaptability to continuous design spaces. Most current models operate under categorical conditioning schemes, which are inadequate for EM applications that often demand conditioning on continuous-valued spectra. Although continuous conditional GANs have been proposed to address such limitations, their application has been largely confined to computer vision domains, with limited exploration in physics-informed design problems such as metasurface synthesis [33]– [35]. Although diffusion models have demonstrated strong generative capabilities, they are predominantly used in computer vision domains

In this regard, the present work proposes a generative inverse design framework for controllable and physically consistent metasurface synthesis under continuous spectral and fabrication constraints. Unlike conventional conditional GANs designed for categorical conditioning, the proposed framework operates directly on continuous EM performance specifications. The model employs a progressively growing Wasserstein GAN with gradient penalty (WGAN-GP) integrated with Feature-wise Linear Modulation (FiLM)-based conditioning, enabling stable propagation of constraints throughout the hierarchical generative process. EM consistency is further embedded directly into training through a surrogate-assisted spectral alignment loss, enabling physics-constrained generation rather than post-generation physics verification. To address the inherent non-uniqueness of inverse EM design, a determinantal point process (DPP)-based diversity regularization strategy is incorporated to generate geometrically distinct yet spectrally consistent metasurface realizations for the same target response, thereby mitigating mode collapse and unstable generation observed in existing approaches. Further, the framework incorporates fabrication-aware constraints by integrating commercially available substrate materials and feasible thickness parameters as input conditions during generation. This ensures that the synthesized metasurface designs remain physically realizable. Unlike most existing inverse design methods that predict the meta-atom geometry under certain fixed assumptions, the proposed framework jointly synthesizes the complete metasurface configuration, including geometry, pattern material, and corresponding resistivity. The efficiency of the developed model has been evaluated using different target reflection characteristics in the frequency range of 2GHz – 18GHz.

The remainder of this paper is organized as follows. Section II presents the theoretical background and architecture of the proposed model. Section III describes the dataset and the training strategies employed. Section IV discusses the results followed by a comprehensive ablation study in Section V. Finally, the conclusions are presented in Section VI.

## II. Proposed Inverse Design Framework

In the proposed framework, inverse design involves the synthesis of the complete configuration of meta-atom from specified reflection characteristics while simultaneously satisfying fabrication constraints. Here, the configuration of a meta-atom is encoded as a three channel image denoted by $x$. The first channel represents the geometric pattern, where a pixel intensity of 255 indicates the presence of metallic or resistive material and 0 denotes its absence. The second channel encodes information on pattern material by weighing the first channel by the normalized resistivity. Finally, the third channel represents the information on number of layers by weighting the first channel with 0 or 1 to represent single-layered and double-layered configurations, respectively. Double-layer configurations are included to model practical scenarios where the metasurface layer requires protection from adverse environmental conditions. The multi-channel colored representation of the meta-atom is shown in Fig. 2. The proposed encoding scheme uniquely captures the design space and effectively reduces the one-to-many mapping enabling stable, physically consistent inverse design. The associated condition, denoted by $y$ consists of the frequency-dependent reflection response represented by the 201-dimensional $|S_{11}|$ over the frequency range of 2–18 GHz, concatenated with the relative permittivity ($\varepsilon_r$), dielectric loss tangent ($tan\ \delta_e$), and thickness ($t$) of the substrate material. i.e.

$$y = [\varepsilon_r, tan\delta_e, t, |S_{11}(f_1)|, |S_{11}(f_2)|, \ldots, |S_{11}(f_n)|] \quad (3)$$

where $|S_{11}(f_i)|$ denotes the absolute value of reflection coefficient at $i^{th}$ frequency point.

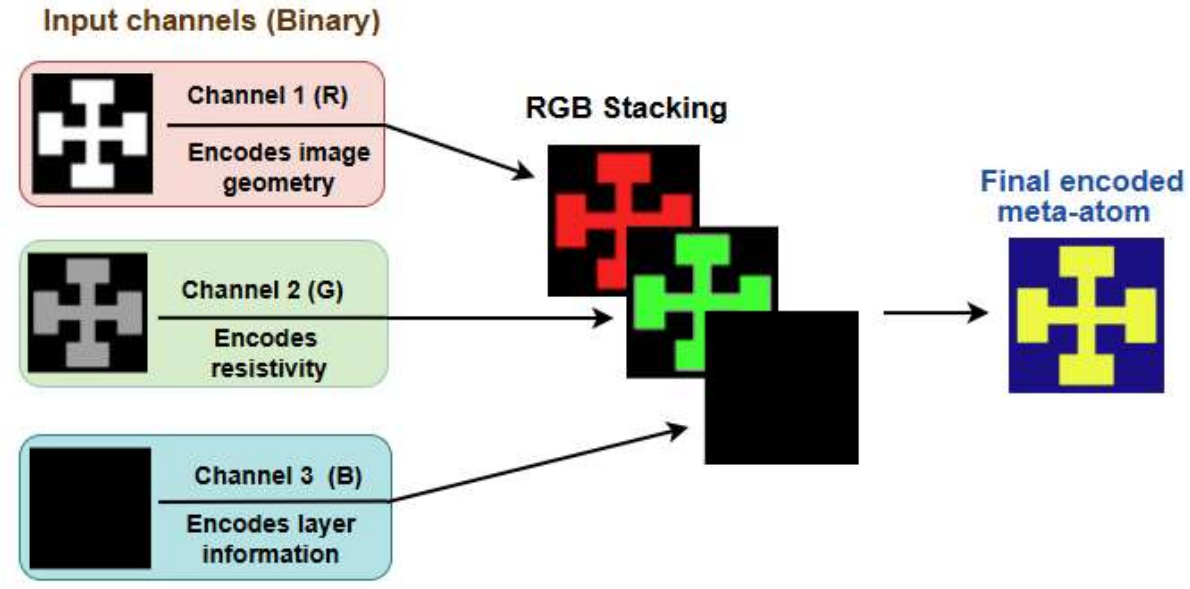

Fig. 2. Encoding scheme used to represent meta-atoms.

In this context, a conditional GAN based framework can be used for the inverse design problem. A typical cGAN consists of two networks: generator ($G$) and discriminator ($D$). The objective of the generator is to synthesize meta atoms, using a specific substrate material, that satisfy the target reflection response. The discriminator aims to distinguish between real and generated designs, while the generator attempts to produce designs indistinguishable from real samples.

To improve training stability and mitigate mode collapse commonly observed in conventional cGANs, a Wasserstein GAN with gradient penalty (WGAN-GP) is employed in the present work. Here, the discriminator is called as a critic that estimates the Wasserstein distance under a Lipschitz continuity constraint, which is enforced via a gradient penalty term. The critic loss in the present case can be expressed as [36],

$$L_C = E_{x \sim p_{data}}[D(x,y)] - E_{z \sim p_z}[D(G(z,y),y)] + \lambda E_{\hat{x} \sim p_{\hat{x}}}[(||\nabla_{\hat{x}} D(\hat{x},y)||_2 - 1)^2] \quad (4)$$

Where $z$, $G(z, y)$, $D(x, y)$ and $\lambda$ denote the noise input to the generator, meta atom generated by $G$ conditioned on $y$, the critic score conditioned on $y$ for real images and gradient penalty coefficient respectively. $p_{data}$ and $p_z$ represent data distribution and noise distribution respectively. $p_{\hat{x}}$ denotes the distribution over points sampled along straight-line interpolations between samples drawn from the real data distribution and the generated data distribution. $\hat{x}$ can be expressed as,

$$\hat{x} = \epsilon x + (1-\epsilon)G(z,y), \epsilon \sim U(0,1) \quad (5)$$

The generator loss can be expressed as,

$$L_G = -E_{z \sim p_z}[D(G(z,y),y)] \quad (6)$$

In the proposed framework, a progressively growing GAN backbone is employed, wherein both the generator and critic are trained by incrementally increasing spatial resolution through the staged addition and smooth blending of layers. Progressively growing GANs stabilize training by learning the data distribution in a coarse-to-fine manner, thereby avoiding the optimization difficulties and mode collapse associated with training directly at full resolution.

While the WGAN-GP objective ensures stable training, it does not explicitly enforce diversity in the generated designs. To address this, a conditioning-driven diversity loss based on determinantal point processes (DPP) [37] is incorporated in the overall loss function. For a batch of generated meta-atom configurations, $B = \{x_1, x_2, \ldots, x_k\}$, a kernel matrix $L^B$ can be constructed as:

$$L_B(u,v) = k(x_u, x_v).\, q(x_u, x_v) \quad (7)$$

Here, $k(x_u, x_v)$ denotes the Gaussian kernel measuring the pairwise similarity between meta-atom geometries based on their feature distance, assigning higher values to similar samples and lower values to dissimilar ones [38]. It can be expressed as,

$$K(x_u, x_v) = e^{\left[-\frac{\|f_u - f_v\|^2}{2\sigma^2}\right]} \quad (8)$$

Where $f_u$ and $f_v$ denote the feature representations of the generated samples drawn from the critic network. σ denotes the bandwidth parameter. $q(x_u, x_v)$ is a performance similarity kernel evaluated as,

$$q(x_u, x_v) = e^{(-\gamma|s(x_u) - s(x_v)|)} \quad (9)$$

$s(x)$ is a conditioning quality score measuring how well a generated design satisfies the target spectral condition. γ controls the trade-off between the conditioning score and diversity. $s(x)$ is computed as the Lambert log exponential transition score (LLETS) [35]. Unlike conventional distance metrics such as $L_1$ and $L_2$, which exhibit weak gradients near optimal solutions and hinder precise conditioning, LLETS provides a stronger self-reinforcing conditioning. It can be expressed as follows:

$$s(x) = LLETS(\varepsilon) = \begin{cases} -\frac{ln\varepsilon}{a};\ \varepsilon > e^{-ae^{W(-\frac{1}{2a})}} \\ e^{-\frac{\varepsilon^2}{2\sigma}}\ ;\ \varepsilon \le e^{-ae^{W(-\frac{1}{2a})}} \end{cases} \quad (10)$$

Where, $\varepsilon$ is the $L_1$ error between desired spectra ($S_{11(desired)}$) and obtained ($S_{11(obtained)}$) spectra. For each generated meta-atom, the corresponding spectrum is predicted using a pre-trained surrogate model plugged into the inverse design framework.

$$\sigma = \frac{e^{-ae^{W(-\frac{1}{2a})}}}{\sqrt{-2W(-\frac{1}{2a})}} \quad (11)$$

where $a$ denote the lambert cutoff that determines the location and quantity of maximum gradient. $W$ denote Lambert $W$ function [39].

The DPP loss for the generator can then be defined as:

$$L_{DPP} = -\frac{1}{|B|} logdet(L_B) \quad (12)$$

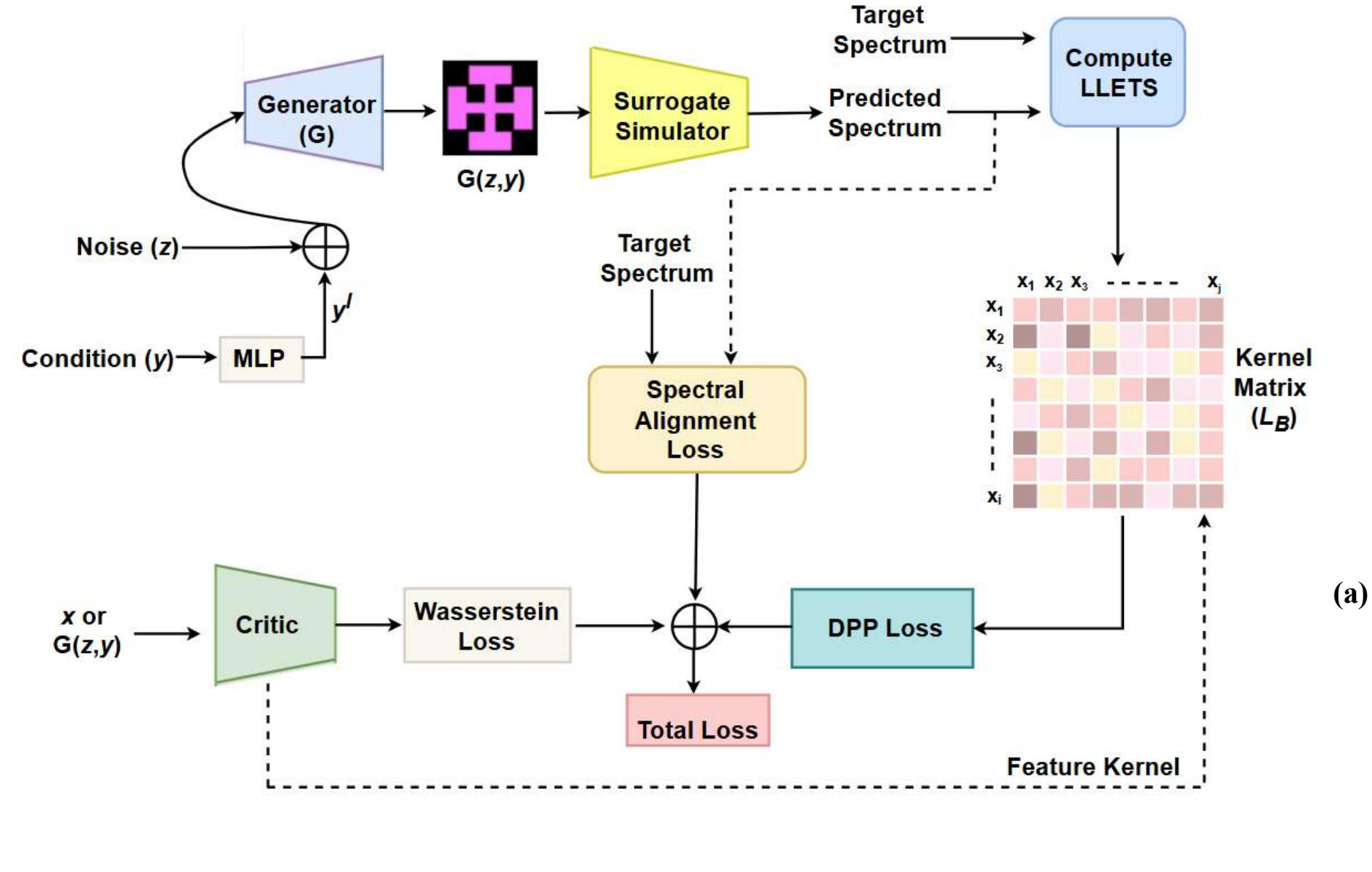

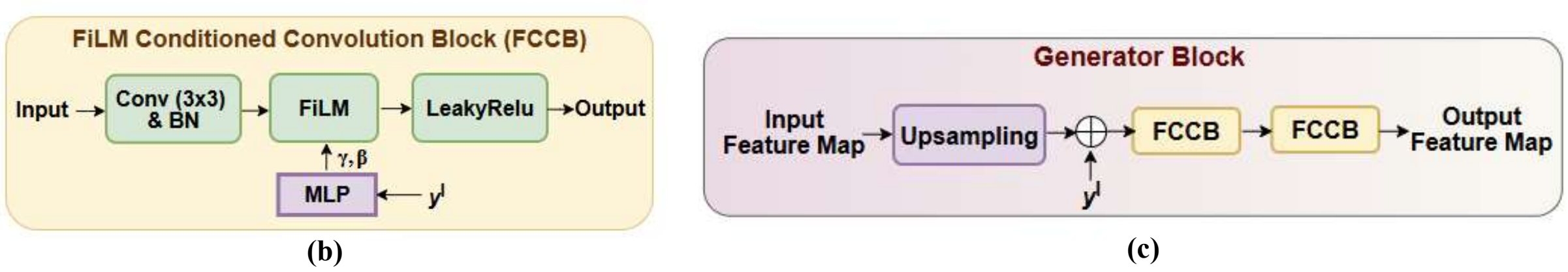

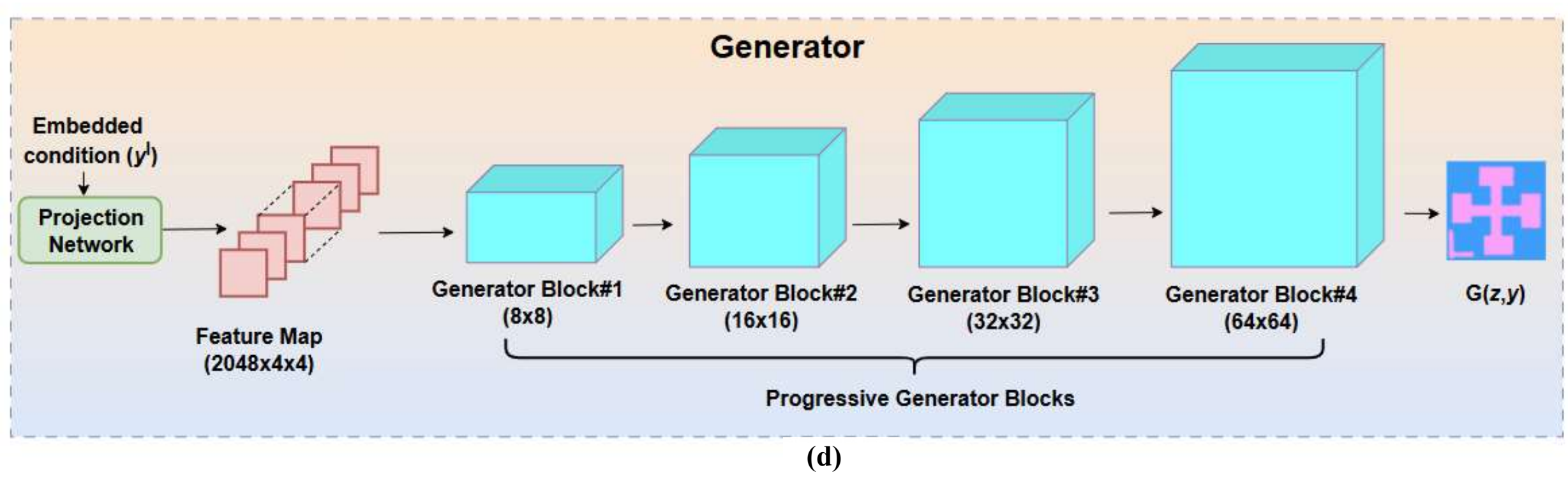

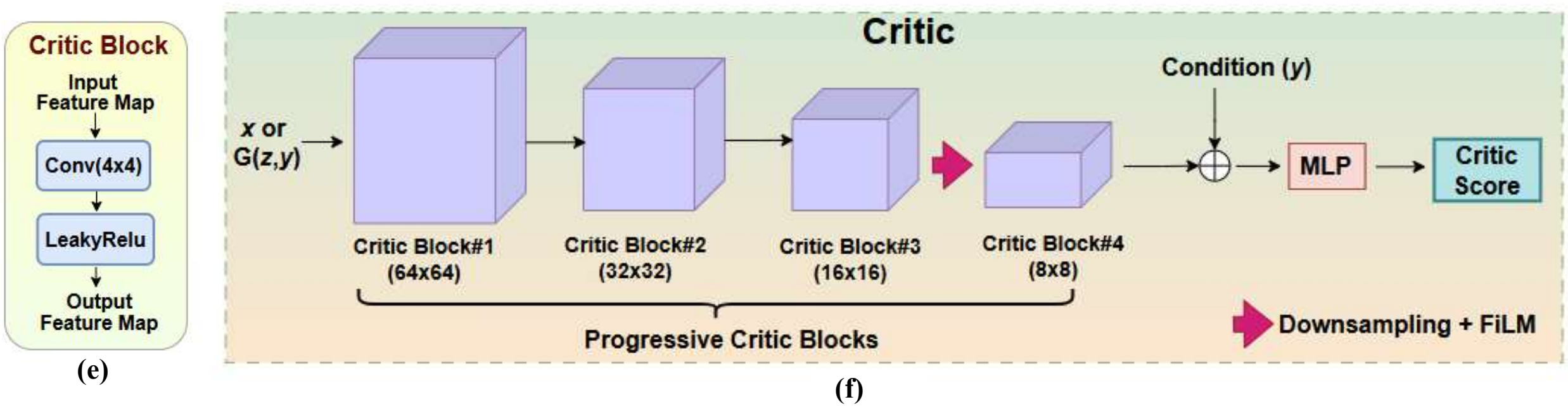

Fig. 3. Architecture of the proposed model (a) Overall block diagram (b) FiLM conditioned convolution block (c) Generator block (d) Generator (e) Critic block (f) Critic.

Maximizing the determinant of $L_B$ encourages the generated samples to be both diverse and conditionally aligned.

To further strengthen physics integration, a spectral alignment loss ($L_{SA}$) is also incorporated into the overall objective as,

$$L_{SA} = \sum_{k=1}^{N_f} \left\| \left|S_{11(desired)}(k)\right| - \left|S_{11(obtained)}(k)\right| \right\|^2 \quad (13)$$

Therefore, the overall loss function of the proposed inverse design framework can be written as,

$$L_{overall}(C, G) = L_C + L_G + L_{DPP} + L_{SA} \quad (14)$$

Another important novelty of the proposed framework is the integration of FiLM-based conditioning [40] which enables effective propagation of continuous spectral and fabrication constraints across the hierarchical stages of the generative process. In this approach, the conditioning vector ($y$), comprising multiple physical attributes such as reflection characteristics and material properties, is first embedded into a meaningful representation $y'$ using a MLP. This embedding is again passed through an MLP to generate the modulation parameters, namely scaling (γ) and shifting (β) coefficients. These parameters are then applied to intermediate feature maps of the network through an affine transformation, enabling feature-wise conditioning across multiple layers. For a feature map $F$, FiLM performs the following transformation,

$$FiLM\ (F) = \gamma \odot F + \beta \quad (15)$$

Where ⊙ denotes element-wise multiplication. This mechanism allows the network to adapt its internal representations based on the target condition.

The overall architecture, incorporating the aforementioned components and conditioning mechanisms, is illustrated in Fig. 3.

## III. Description of Dataset and Training Strategies

A diverse dataset spanning geometries, materials, and reflection spectra is critical for learning robust mappings between meta-atoms and their EM responses. The dataset generated for the present study consists of 39664 data points covering 30 different types of meta-atoms. Each meta-atom variant has been simulated using appropriate parameter combinations, commercially available substrate materials with feasible thicknesses, different pattern materials (metal/resistive) and in different configurations (single-layer/double-layer). This has been done so that the model will have sufficient data to learn the variations in reflection spectra arising from change in geometry, structure and material properties. The geometrical parameters considering fabrication tolerances for each variant have also been chosen after a rigorous parametric analysis so as to generate a wide range of reflection characteristics in the frequency range of 2-18 GHz**.** The configurations of meta-atoms considered in the dataset are shown in Fig. 4. The data points have been obtained through full-wave EM simulations performed in CST Studio Suite [41]. The analysis of the meta-atoms has been carried out using frequency-domain solver under periodic boundary conditions in the lateral directions, open (add space) boundaries along the normal direction, and plane-wave excitation.

Once the data has been generated, to mitigate bias arising from dominant spectral patterns, the dataset has been organized into different response categories like narrow-band, wideband, ultra-wideband, etc. During training of the model, a weighted sampling strategy has been then employed wherein sampling probabilities are assigned inversely to the size of each spectral category. This ensures higher sampling probability to under-represented spectral categories thereby promoting balanced learning. Further, the dataset has been divided into train, validation and test sets using a stratified category-wise strategy with adaptive allocation. While well-represented categories use an 80:10:10 split, underrepresented ones have been partitioned using a minimum-count approach to maintain evaluation coverage without compromising training data.

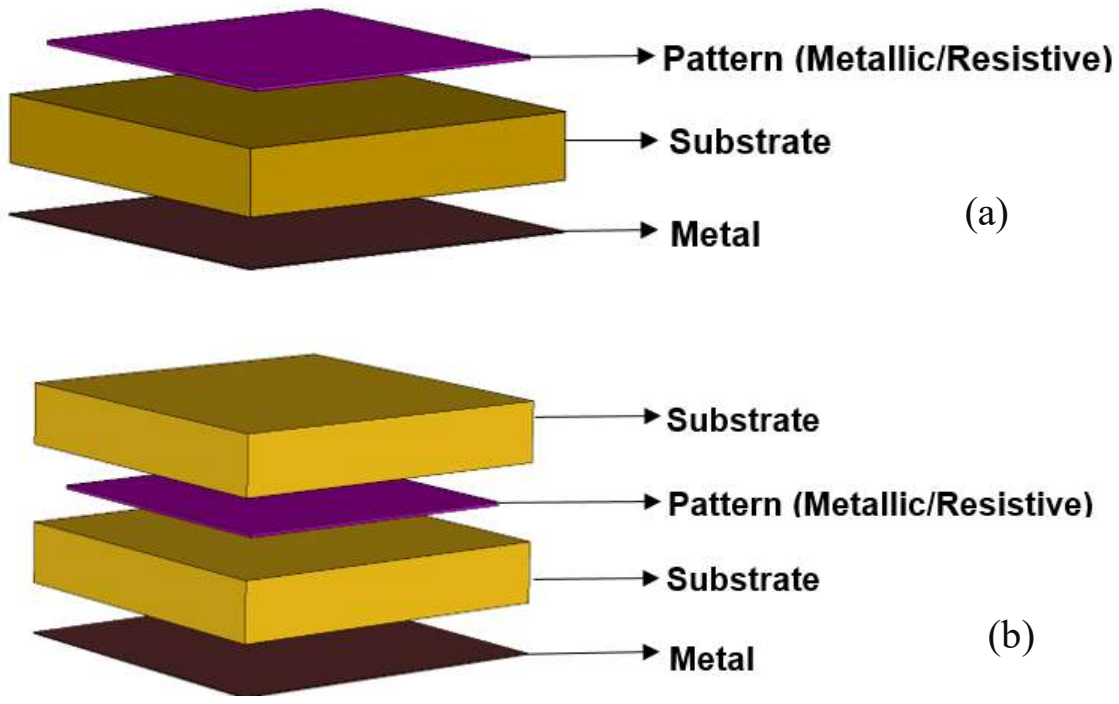

Fig. 4. Configuration of metasurface based RAS models included in the dataset (a) Single-layered RAS (b) Embedded RAS.

## IV. Results and Discussion

In this section, the performance of the proposed framework is evaluated through qualitative and quantitative comparisons between the desired and generated EM responses. The model has been initially optimized through systematic hyperparameter tuning. The model with the optimized set of hyper parameters has been trained for 280000 training steps using AdamW optimizer in conjunction with a cosine annealing learning rate schedule (Initial learning rate of $G$ and $D$ = 1e-4). The training and validation losses converged within 20,000 training steps. The trained model has been subsequently deployed in inference mode to generate meta-atoms for given conditions, using samples in test dataset. Fig. 5 compares the desired spectra with those obtained from full-wave simulations of the generated meta-atoms in CST Studio Suite.

The close agreement between the target and generated spectra demonstrates the effectiveness of the proposed conditioning framework, wherein FiLM-based modulation and spectral alignment loss collectively enable accurate physics-constrained generation under continuous spectral conditions. Furthermore, the synthesized meta-atom designs exhibit structurally realistic and fabrication-feasible features, highlighting the effectiveness of the dataset construction methodology. The generated images of meta-atoms are shown as insets in Fig. 5.

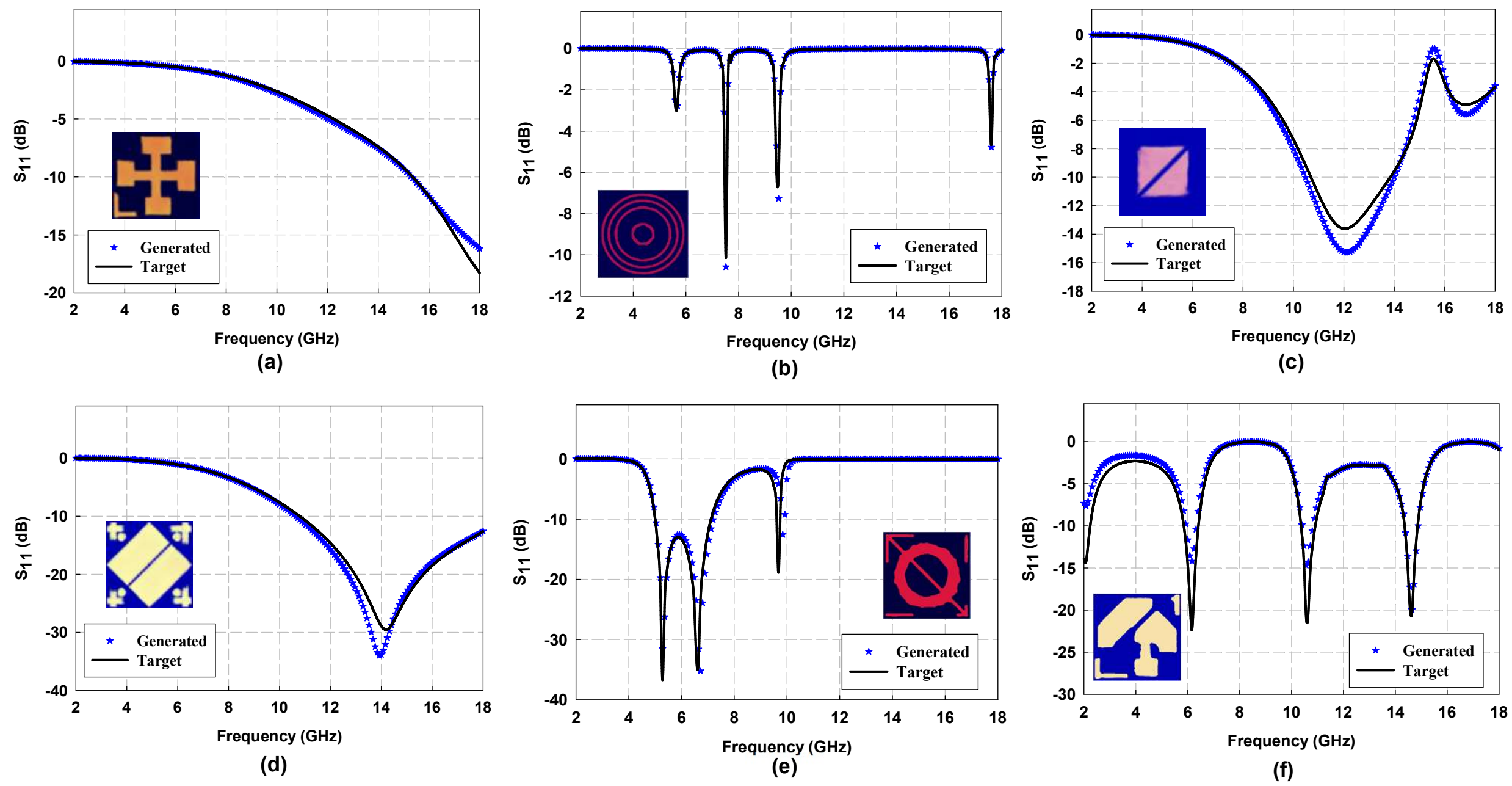

Fig. 5. Comparison between input spectrum and spectra obtained on simulation of generated design. (a) Case-1 (Material: AD1000 ($\varepsilon_r$=10.2, $tan\delta$=0.0023, $t$=1.27mm); Single-layered; Resistive pattern (59 ohm/sq.) pattern) (b) Case-2 (Material: Kappa438 ($\varepsilon_r$=4.38, $tan\delta$=0.005, $t$=1.016mm); Single-layered; Metallic pattern) (c) Case-3 (Material: RO3206 ($\varepsilon_r$=6.6, $tan\delta$=0.0027, $t$=1.27mm); Double-layered; Resistive pattern (62 ohm/sq.)) (d) Case-4 (Material: RO4835 ($\varepsilon_r$=3.48, $tan\delta$=0.0037, $t$=1.524mm); Double-layered; Resistive (97 ohm/sq.) pattern) (e) Case-5 (Material: AD350A ($\varepsilon_r$=3.54, $tan\delta$=0.0033, $t$=3.048mm); Single-layered; Metallic Pattern) (f) Case-6 (Material: TMM13i ($\varepsilon_r$=12.2, $tan\delta$=0.0019, $t$=5.08mm); Double-layered; Resistive(91 ohm/sq.) pattern.).

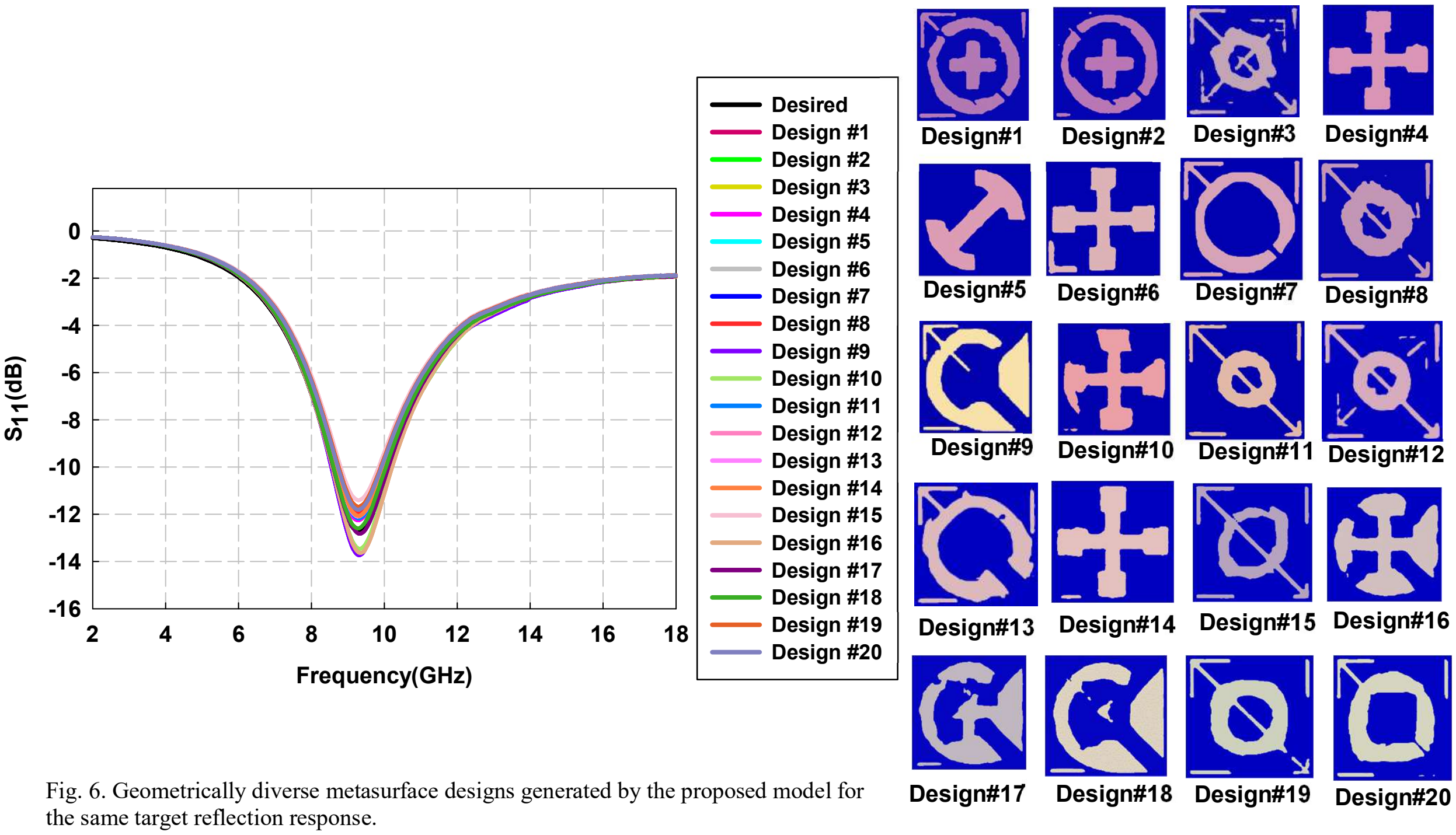

Fig. 6. Geometrically diverse metasurface designs generated by the proposed model for the same target reflection response.

Fig. 5 demonstrates that the proposed model is capable of generating meta-atoms satisfying diverse EM constraints, including multi-resonant, narrowband, and wideband reflection characteristics. Moreover, the model exhibits strong feature-level generalization by synthesizing hybrid meta-atom geometries (Fig. 5a and Fig. 5f) through effective interpolation and fusion of structural features learned from multiple meta-atom patterns in the training dataset.

The impact of DPP loss in enhancing diversity has been evaluated by generating multiple geometries for the same condition, as shown in Fig. 6. The model produces several structurally realistic designs yielding similar reflection spectra, demonstrating its ability to capture diverse solutions, offering flexibility for practical realization.

**Table I** Summary of ablation analysis

| Model Configuration | MSE | AAE | D | BAA | Valid EM Designs (%) |
|---|---|---|---|---|---|
| Conventional WGAN-GP | 0.0894 | 0.1932 | 0.3634 | 0.2150 | 21.05 |
| Progressively growing WGAN-GP | 0.0766 | 0.1775 | 0.3959 | 0.3162 | 31.58 |
| Progressively growing WGAN-GP with Surrogate Model | 0.0525 | 0.1373 | 0.5307 | 0.4457 | 42.11 |
| Progressively growing WGAN-GP with Surrogate Model and DPP loss | 0.0391 | 0.1155 | 0.7274 | 0.4837 | 63.16 |
| **Progressively growing WGAN-GP with Surrogate model, DPP loss and FiLM conditioning (Proposed)** | **0.0052** | **0.0343** | **0.8730** | **0.8533** | **89.57** |

## V. Ablation Analysis

This section presents an ablation analysis to systematically demonstrate the progressive enhancement in the performance of the model achieved through the incremental integration of advanced architectural and conditioning components, including progressively growing network architecture, surrogate model-based spectral guidance, DPP-based diversity regularization, and FiLM-based conditioning mechanisms into the baseline WGAN-GP framework. The ablation analysis provides important insights into the influence of each modification on key evaluation metrics [42] like average spectral mean squared error (MSE), accumulated average error (AAE), diversity metric ($D$), average band alignment accuracy (BAA), and the percentage of valid designs. All evaluation metrics are computed over the test dataset. They are defined as:

$$MSE = \frac{1}{N_{test}} \sum_{j=1}^{N_{test}} \frac{1}{N_f} \sum_{i=1}^{N_f} \left( |S_{11(gen)}(f_i)| - |S_{11(target)}(f_i)| \right)^2 \quad (16)$$

$$AAE = \frac{1}{N_{test}} \sum_{j=1}^{N_{test}} \frac{1}{N_f} \sum_{i=1}^{N_f} \left| |S_{11(gen)}(f_i)| - |S_{11(target)}(f_i)| \right| \quad (17)$$

Where $|S_{11(gen)}(f_i)|$ and $|S_{11(target)}(f_i)|$ denote the generated and target spectral response at $i^{th}$ frequency point ($f_i$). $N_f$ denote total number of frequency points and $N_{test}$ denote number of data points in the test data set.

To quantify geometric diversity, the following average pairwise structural dissimilarity metric ($D$) is employed: [43]:

$$D = \frac{2}{N(N-1)} \sum_{i<j} \left[ \lambda d_1 + (1-\lambda) d_2 \right] \quad (18)$$

Where, $d_1 = 1 - \frac{|M_i \cap M_j|}{|M_i \cup M_j| + \epsilon}$; $d_2 = 1 - \frac{|R_i^{(\delta)} \cap R_j^{(\delta)}|}{|R_i^{(\delta)} \cap R_j^{(\delta)}| + \epsilon}$

$M_i$ and $M_j$ denote the binary masks corresponding to the $i^{th}$ and $j^{th}$ generated meta-atom geometries, respectively, where 1 represents the presence of pattern material and 0 represents the absence of it. $d_1$ measures the variations in pattern material distribution between a pair of images using the intersection over union (IoU) distance between binary masks. $R_i^{(\delta)}$ and $R_j^{(\delta)}$ denote the boundary regions of the corresponding masks, obtained by selecting pixels lying within a distance $\delta$ from the metal edges. $d_2$ quantifies fine-scale geometric differences by comparing the contour regions. The weighting factor $\lambda$ controls the relative contribution of global structural variation and local boundary variation. A small constant $\epsilon$ is introduced in the denominator to ensure numerical stability. $N$ denotes the number of generated images.

BAA measures the overlap between the desired and generated absorption regions and is defined as,

$$BAA = \frac{1}{N_{test}} \sum_{j=1}^{N_{test}} \frac{|B_{target}(j) \cap B_{gen}(j)|}{|B_{target}(j)|} \quad (19)$$

where $B_{target}$ and $B_{gen}$ refer to the set of frequency points in the target spectrum and generated spectrum respectively where $|S_{11}| \leq -10$dB. The percentage of valid designs denote the fraction of generated designs that has a band alignment greater than 0.8.

Table I presents the summary of the ablation analysis *w.r.t.* the above-mentioned evaluation metrics. It is to be noted that, for the first four models listed in Table I, conditioning is performed by concatenating the condition vector with the latent noise input to the generator. In the critic, the output feature vector from the final critic block is flattened and combined with the condition vector prior to the final prediction layer. Table I clearly demonstrates that the proposed enhancements collectively enable the generation of structurally diverse, high-fidelity, and electromagnetically consistent metasurface configurations with significantly improved conformity to the target reflection spectra. Furthermore, the proposed framework generates a suitable meta-atom design for a specified target response in few seconds, whereas conventional iterative full-wave simulation-driven approaches may require several months under comparable computational resources.

## VI. Conclusion

This paper presented a physics-aware generative framework for inverse design of metasurface-based RAS. The proposed progressively growing WGAN-GP with FiLM-based conditioning enabled stable propagation of continuous EM constraints across the model, while a surrogate-assisted spectral alignment loss enforced physics-constrained generation during training. In addition, a DPP-based diversity regularization strategy enabled generation of geometrically distinct yet spectrally consistent metasurface designs for identical target responses, addressing the inherent non-uniqueness of inverse design. In addition, fabrication-aware constraints involving material properties and feasible thicknesses were incorporated to ensure practical realizability. Simulation studies demonstrated accurate synthesis of narrowband, multi-resonant, and broadband absorber configurations in the 2–18 GHz frequency range with strong agreement between target and simulated responses. The final proposed framework achieved an average mean squared error of 0.0052, accumulated average error of 0.0343, diversity score of 0.8730, band alignment accuracy of 0.8533, and a valid EM design generation rate of 89.57%, clearly demonstrating its capability to generate highly accurate, diverse, and electromagnetically consistent metasurface configurations. Overall, the proposed framework provides a scalable and computationally efficient pathway for rapid inverse design of practically realizable metasurfaces for stealth applications. Future extensions of the proposed framework can focus on incorporating oblique incidence and polarization-dependent EM responses to improve applicability to realistic operating conditions.

## ACKNOWLEDGMENT

We express our gratitude to Council of Scientific and Industrial Research (CSIR), India for supporting this research activity. We also express our gratitude to CSIR-4PI for providing access to HPC facility.

## DATA AVAILABILITY STATEMENT

The dataset supporting this study is not publicly available due to confidentiality restrictions associated with sponsored research projects involving proprietary metasurface designs.

## CONFLICT OF INTEREST

The authors declare that they have no conflict of interest.

## REFERENCES

[1] C. L. Holloway, E. F. Kuester, J. A. Gordon, J. O'Hara, J. Booth, and D. R. Smith, "An overview of the theory and applications of metasurfaces: The two-dimensional equivalents of metamaterials," *IEEE Antennas Propag. Mag.*, vol. 54, no. 2, pp. 10–35, Apr. 2012

[2] H.-T. Chen, A. J. Taylor, and N. Yu, "A review of metasurfaces: physics and applications," *Rep. Prog. Phys.*, vol. 79, no. 7, p. 076401, 2016.

[3] J. B. Pendry, "Negative refraction makes a perfect lens," *Phys. Rev. Lett.*, vol. 85, no. 18, pp. 3966–3969, Oct. 2000.

[4] D. R. Smith, W. J. Padilla, D. C. Vier, S. C. Nemat-Nasser, and S. Schultz, "Composite medium with simultaneously negative permeability and permittivity," *Phys. Rev. Lett.*, vol. 84, no. 18, pp. 4184–4187, May 2000.

[5] S. B. Glybovski, S. A. Tretyakov, P. A. Belov, Y. S. Kivshar, and C. R. Simovski, "Metasurfaces: From microwaves to visible," *Physics Reports*, vol. 634, pp. 1–72, 2016.

[6] N. Yu, P. Genevet, M. A. Kats, F. Aieta, J.-P. Tetienne, F. Capasso, and Z. Gaburro, "Light propagation with phase discontinuities: Generalized laws of reflection and refraction," *Science*, vol. 334, no. 6054, pp. 333–337, Oct. 2011, doi: 10.1126/science.1210713.

[7] A. V. Kildishev, A. Boltasseva, and V. M. Shalaev, "Planar photonics with metasurfaces," *Science*, vol. 339, no. 6125, p. 1232009, Mar. 2013, doi: 10.1126/science.1232009.

[8] N. Yu and F. Capasso, "Flat optics with designer metasurfaces," *Nature Materials*, vol. 13, no. 2, pp. 139–150, Feb. 2014, doi: 10.1038/nmat3839.

[9] X. Ni, A. V. Kildishev, and V. M. Shalaev, "Metasurface holograms for visible light," *Nature Communications*, vol. 4, p. 2807, Dec. 2013, doi: 10.1038/ncomms3807.

[10] M. Khorasaninejad, W. T. Chen, R. C. Devlin, J. Oh, A. Y. Zhu, and F. Capasso, "Metalenses at visible wavelengths: Diffraction-limited focusing and subwavelength resolution imaging," *Science*, vol. 352, no. 6290, pp. 1190–1194, Jun. 2016, doi: 10.1126/science.aaf6644.

[11] M. Khorasaninejad and F. Capasso, "Metalenses: Versatile multifunctional photonic components," *Science*, vol. 358, no. 6367, p. eaam8100, Nov. 2017, doi: 10.1126/science.aam8100.

[12] C. Pfeiffer and A. Grbic, "Metamaterial Huygens' surfaces: Tailoring wave fronts with reflectionless sheets," *Physical Review Letters*, vol. 110, no. 19, p. 197401, May 2013, doi: 10.1103/PhysRevLett.110.197401.

[13] F. Ding, A. Pors, and S. I. Bozhevolnyi, "Gradient metasurfaces: A review of fundamentals and applications," *Reports on Progress in Physics*, vol. 81, no. 2, p. 026401, Feb. 2018, doi: 10.1088/1361-6633/aa8732.

[14] S. Jahani and Z. Jacob, "All-dielectric metamaterials," *Nature Nanotechnology*, vol. 11, no. 1, pp. 23–36, Jan. 2016, doi: 10.1038/nnano.2015.304.

[15] F. A. A. Nugroho, G. W. Castellanos, P. Bai, I. Darmadi, J. Fritzsche, J. Gómez Rivas, and A. Baldi, "Inverse designed plasmonic metasurface with parts per billion optical hydrogen detection," *Nature Materials*, vol. 18, no. 5, pp. 489–495, May 2019, doi: 10.1038/s41563-019-0296-9.

[16] V. Joy, A. Dileep, P. V. Abhilash, R. U. Nair, H. Singh, "Metasurfaces for stealth applications: A comprehensive review," *Journal of Electronics Materials*, vol. 50, no. 6, pp. 3129–3148, Jun. 2021.

[17] M. Khorasaninejad and F. Capasso, "Broadband multifunctional efficient meta-gratings based on dielectric waveguide phase shifters," *Nano Letters*, vol. 15, no. 10, pp. 6709–6715, Oct. 2015.

[18] O. Bouvard, M. Lanini, and L. Burnier, "Structured transparent low emissivity coatings with high microwave transmission," *Applied Physics A*, vol. 123, no. 1, Art. no. 66, 2017.

[19] Z. Li, R. Pestourie, Z. Lin, S. G. Johnson, and F. Capasso, "Empowering metasurfaces with inverse design: Principles and applications," *ACS Photonics*, vol. 9, no. 7, pp. 2178–2192, Jul. 2022, doi: 10.1021/acsphotonics.1c01850.

[20] M. M. R. Elsawy, S. Lanteri, R. Duvigneau, G. Brière, M. S. Mohamed & P. Genevet, "Global optimization of metasurface designs using statistical learning methods," *Scientific Reports*, vol. 9, p. 17918, 2019, doi: 10.1038/s41598-019-53878-9.

[21] C. Kang, C. Park, M. Lee, J. Kang, M. S. Jang, and H. Chung, "Large-scale photonic inverse design: Computational challenges and breakthroughs," *Nanophotonics*, vol. 13, no. 20, pp. 3765–3792, Jun. 2024, doi: 10.1515/nanoph-2024-0127.

[22] S. So, T. Badloe, J. Noh, J. Bravo-Abad, and J. Rho, "Deep learning enabled inverse design in nanophotonics," *Nanophotonics*, vol. 9, no. 5, pp. 1041–1057, 2020.

[23] D. Liu, Y. Tan, E. Khoram, and Z. Yu, "Training deep neural networks for the inverse design of nanophotonic structures," *ACS Photonics*, vol. 5, no. 4, pp. 1365–1369, 2018.

[24] Z. Liu, D. Zhu, K. Lee, A. S. Kim, and J. R. W. Mark, "Generative model for the inverse design of metasurfaces," ACS Nano, vol. 12, no. 7, pp. 7073–7083, 2018.

[25] A. Nezaratizadeh, S. M. Hashemi, and M. Bod., "Prediction of multi-layer metasurface design using conditional deep convolutional generative adversarial networks," International Journal for Light and Electron Optics, vol. 313, p. 172005, 2024.

[26] G. Dai, H. Li, X. Zhang, and Y. Liu, "Single-layer metasurface design with symmetrical free-form patterns using generative adversarial networks," Applied Soft Computing, vol. 127, Art. no. 109353, 2022.

[27] A. Mall, A. Patil, A. Sethi & A. Kumar, "A cyclical deep learning based framework for simultaneous inverse and forward design of nanophotonic metasurfaces," *Scientific Reports*, Article number: 19427, 2020.

[28] A. Mall, A. Patil, D. Tamboli, A. Sethi and A. Kumar, "Fast design of plasmonic metasurfaces enabled by deep learning," *Journal of Physics D: Applied Physics*, vol. 53, no. 49, October 2020.

[29] P. Naseri and S. V. Hum, “A generative machine learning-based approach for inverse design of multilayer metasurfaces,” IEEE Transactions on Antennas and Propagation, vol. 69, no. 9, pp. 5725 5739, Sep. 2021.
[30] Q. Li, J. Wang, T. Lei, T. Xiang, C. Qin and M. Yang, “Design of Metamaterials for Absorbers Based on Variational Autoencoder,” IEEE Access, vol. 12, pp. 92328-92336, 2024.
[31] W. Lin, J. Zhang, Z. Zou, Y. Lin, Y. Peng, W. Yang, Y. Zhang, T. Guo, C. Wu, and X. Zhou, “VAE enhanced tandem neural network for reverse design of metasurface structural-colors with high efficiency and accuracy,” Optics Communications, vol. 601, p. 132760, 2026.
[32] X. Zheng, J. Shiomi and T. Yamada, “Optimizing metamaterial inverse design with 3D conditional diffusion model and data augmentation,” Advanced Materials Tech., vol. 10, 10p., 2025.
[33] W. Chen and F. Ahmed, “PaDGAN: A generative adversarial network for performance augmented diverse designs,” arXiv preprint arXiv:2002.11304, 2020.
[34] X. Ding, Y. Wang, Z. Xu, W. J. Welch, and Z. J. Wang, “Continuous conditional generative adversarial networks: Novel empirical losses and label input mechanisms,” IEEE Trans.Pattern Analysis and Machine Intelligence, vol. 45, no. 7, pp. 8143-8158, Jul. 2023.
[35] A. H. Nobari, W. Chen, and F. Ahmed, “PcDGAN: A Continuous Conditional Diverse Generative Adversarial Network for Inverse Design,” in *Proc. 27th ACM SIGKDD Conf. Knowl. Discovery Data Mining (KDD '21)*, Virtual Event, Singapore, Aug. 2021, pp. 1–12. doi: 10.1145/3447548.3467414.
[36] I. Gulrajani, F. Ahmed, M. Arjovsky, V. Dumoulin, and A. C. Courville, “Improved training of Wasserstein GANs,” in *Adv. Neural Inf. Process. Syst. (NIPS)*, vol. 30, pp. 5767–5777, 2017.
[37] A. Kulesza and B. Taskar, *Determinantal Point Processes for Machine Learning*. Hanover, MA, USA: Now Publishers Inc., Nov. 2012, pp. 1–178, ISBN: 978-1-60198-628-3.
[38] T. Hofmann, B. Schölkopf, and A. J. Smola, “Kernel methods in machine learning,” The *Annals of Statistics*, vol. 36, no. 3, pp. 1171–1220, 2008.
[39] R. M. Corless, G. H. Gonnet, D. E. G. Hare, D. J. Jeffrey, and D. E. Knuth, “On the Lambert W function,” *Advances in Computational Mathematics*, vol. 5, no. 1, pp. 329–359, 1996, doi: 10.1007/BF02124750.
[40] E. Perez, F. Strub, H. de Vries, V. Dumoulin, and A. Courville, “FiLM: Visual reasoning with a general conditioning layer,” in Proceedings of the Thirty-Second AAAI Conference on Artificial Intelligence, Article No.: 483, pp. 3942 - 3951, 2018.
[41] CST Studio Suite®, CST AG, Germany, www.cst.com
[42] C. M. Bishop, *Pattern Recognition and Machine Learning*. New York, NY, USA: Springer, 2006.
[43] B. Cheng, R. Girshick, P. Dollár, A. C. Berg, and A. Kirillov, “Boundary IoU: Improving object-centric image segmentation evaluation,” in *Proc. IEEE/CVF Conf. Comput. Vis. Pattern Recognit. (CVPR)*, Nashville, TN, USA, 2021, pp. 15334–15342.